\documentclass[%
 reprint,
 showkeys,
nofootinbib,
 amsmath,amssymb,
 aps,
]{revtex4-2}

\usepackage{graphicx}
\usepackage{dcolumn}
\usepackage{bm}
\usepackage{natbib}
\def\gsim{\mathrel{\raise.5ex\hbox{$>$}\mkern-14mu
             \lower0.6ex\hbox{$\sim$}}}
\def\lsim{\mathrel{\raise.3ex\hbox{$<$}\mkern-14mu
             \lower0.6ex\hbox{$\sim$}}}

\begin{document}

\preprint{ }

\title{Interference with non-interacting free particles and a special type of detector}


\author{Ioannis Contopoulos$^{1}$}
 \email{icontop@academyofathens.gr}
\author{Athanasios C. Tzemos$^{1}$}
 \email{atzemos@academyofathens.gr\ (corresponding author)}
\author{Foivos Zanias$^{2}$}
\author{George Contopoulos$^{1}$}
\affiliation{$^1$Research Center for Astronomy and Applied Mathematics, Academy of Athens, GR 11527 Athens, Greece\\
$^2$Department of Physics, National and Kapodistrian University of Athens, GR 15783 Zografos, Greece}


\date{\today}

\begin{abstract}
It is shown how a classical detector that collects 
non-interacting individual classical massive free particles 
can generate a quantum interference pattern. The proposed 
classical picture requires that particles 
carry the information of a phase equal to an action 
integral along their trajectory. At the point of their 
detection, a special type of detector collects the phases 
from all individual particles reaching it, adds them up 
over time as complex numbers, and divides them by the 
square root of their number. The detector announces a 
number of detections equal to the square of the amplitude 
of the resulting complex number. An interference pattern 
is gradually built from the collection of particle 
phases in the detection bins of the detector after several 
repetitions of the experiment. We obtain perfect agreement 
with three solutions of the Schr\"{o}dinger equation for 
free particles: a Gaussian wavepacket, two Gaussian wavepackets 
approaching each other, and a Gaussian wavepacket reflecting off a wall.
The main conclusion of the present work is that, if there 
are macroscopic detectors operating as proposed, then the 
interference of quantum mechanics is basically due to 
the detectors that collect the particles. 
Finally, a simple physical experiment with a single-photon 
detector is proposed that will be able to test our theory.

\end{abstract}

\keywords{computer simulation, quantum theory, quantum interpretations}
\maketitle

\section{Recorded vs Detected Particles}
		
Quantum mechanics has a multitude of interpretations (Copenhagen,  de Broglie-Bohm, von Neumann-Wigner, stochastic mechanics, many worlds, etc.) \citep{harrigan2010einstein,bohn1952suggested,nelson1966derivation,nelson1985quantum,holland1995quantum,
maudlin2018foundations,nelson2012review,merzbacher1998quantum,	styer2002nine,bassi2013models,bacciagaluppi2009quantum, everett1957relative}. 
There are still problems with the ontological duality that microscopic particles, whenever detected, manifest themselves as independent particles, but when a large number of them is detected, their distribution manifests wave-like characteristics like interference. Feynman famously quoted that wave-like characteristics are ‘impossible, absolutely impossible to explain in any classical
way and have in them the heart of quantum mechanics’
(Feynman Lectures on Physics \cite{4}). We, as several others before us \citep{bohn1952suggested,nelson1966derivation,jin2010corpuscular}, propose that particles are indeed independent individual entities, and there must be another way to interpret the wave-like characteristics of their distributions. An interpretation of quantum mechanics without the conceptual problem of particle-wave duality may be that individual particles carry some type of information in the form of a phase that is collected by a special type of detector who manifests wave-like characteristics only after processing a large number of particles.

In the present work we investigate a model of this special detector in the case of non-relativistic free particles of mass $m$. What is proposed is that individual particles carry a phase equal to their action integral
\begin{eqnarray}
\phi &= & \int\frac{\pi {\rm d}l}{\lambda}+\phi_o\nonumber\\
&\equiv& \int\frac{\frac{1}{2}mv^2}{\hbar}{\rm d}t+\phi_o\ ,
\label{phi}
\end{eqnarray}
where the ${\rm d}l$ integration is taken along the particle trajectory, and $\lambda$ is a characteristic length equal to $\lambda=h/(m\ {\rm d}l/{\rm d}t)$, and $\phi_o$ is some initial particle phase. Here, $h$ is the Planck constant, $\hbar\equiv h/(2\pi)$, $m$ is the particle mass, and $v\equiv {\rm d}l/{\rm d}t$ is the particle velocity.

Our special detector collects the $N$ particles that reach its detection bin around the position $({\bf x};t)$ after a large number of repetitions of the experiment and processes their collected phases $\phi_j$ by calculating the expression
\begin{equation}
\overline{\Psi}({\bf x};t)\equiv
\sum\nolimits_{k=1}^K
\frac{1}{\sqrt{N_k}}
\sum\nolimits_{j=1}^{N_k} {\rm e}^{i\phi_j}\ .
\label{Psi}
\end{equation}
It is assumed here that particles reach the particular detection position $({\bf x};t)$ from $K$ different directions, and $\phi_j$ are the corresponding phases along each trajectory. $\sum\nolimits_{k=1}^K N_k=N$. Finally, it is assumed that all $N$ particles that reach the detector are {\it detected}, but 
\begin{equation}
\overline{N}\equiv |\overline{\Psi}|^2\equiv\left|
\sum\nolimits_{k=1}^K
\frac{1}{\sqrt{N_k}}
\sum\nolimits_{j=1}^{N_k} {\rm e}^{i\phi_j}\right|^2
\label{Nrecorded}
\end{equation}
particles are supposed to be {\it recorded}. Depending on the particular phases of the particles that reach the detector, $\overline{N}$ may be smaller, equal, or greater than $N$. In general $\overline{N}\neq N$. 
{
Notice that what is important in this procedure is the phase difference with respect to the initial phase value in eq.~(\ref{phi}). There is no detector that can directly measure phases. However, there are detectors that can measure phase differences (e.g. detectors that use the Aharonov-Bohm effect on charged particles moving in a non-vanishing magnetic vector potential or a step-wise electric potential).

}

In a world where all measurements are performed according to our model of this special type of local detector, the wavefunction is not something that characterizes each individual particle. Particles have no a-priori knowledge of their distribution in space and time $N({\bf x};t)$. This is obtained {\it after} a large number of repetitions of the experiment. This is in contrast to other non-local particle theories like Bohmian mechanics \citep{bohn1952suggested,holland1995quantum}. The particles do not know anything about the wavefunction, nor about their distributions $N({\bf x};t)$ or $\overline{N}({\bf x};t)$. In order to determine their distribution after several repetitions of the experiment, detectors must be placed at various positions in space and time to collect detections over and over again. The role of the detector is to collect the phases of the particles that reach it, and process them according to eq. (\ref{Psi}). Then one can gradually build a spatial distribution $\overline{N}({\bf x};t)$ following eq.~(\ref{Nrecorded}). In order for wave-like interference effects to manifest themselves, it is important that the size of the detection bins is smaller than the length $\lambda$. 

{ Our formulation is very similar to the {\it Initial Value Representation} (hereafter IVR) method proposed by Miller in the early 1970s and largely improved since  \cite{Miller1970,Miller2001,Miller2012}, which is itself based on Feynman's path integrals \citep{Feynman1965}. However, in the present work we focus on the way real physical detectors operate (this might be seen as a possible physical realization of the IVR
method in Nature). In fact, as will be seen below, one can test whether real detectors operate as proposed with a simple physical experiment that we are currently planning to perform. That experiment will test whether the number $\overline{N}$ of particles {\it recorded} by the detector differs from the number $N$ of {\it detected} particles. If that is indeed the case, one will conclude that the interference of quantum mechanics is due to the detectors that collect the particles. This remains to be confirmed.}

We begin in \S~II by investigating a Gaussian distribution of particles moving in one dimension. We { consider} a large number of detectors with finite detection volumes that collect and process the particle phases every time a particle passes through them. The initial conditions and the parameters of our model are adjusted in order to reproduce the solution of the Schr\"{o}dinger equation for a free Gaussian wavepacket. An analytic approximation of our result is also obtained.
In \S~III a second Gaussian distribution of particles is introduced in phase with the first one that moves towards the first one and repeat the above calculation. A clear interference pattern emerges after a large enough number of independent repetitions of the experiment, in excellent agreement with the solution of the Schr\"{o}dinger equation. In \S~IV the above are repeated for a Gaussian distribution reflecting off a wall and again interference is found, in excellent agreement with the solution of the Schr\"{o}dinger equation. 
We conclude with a summary of our results in \S~V, and discuss briefly one case of photon interference in the Appendix.

\section{The Gaussian free-particle wavepacket}

\begin{figure*}
 \centering
 \includegraphics[width=1.05\textwidth]{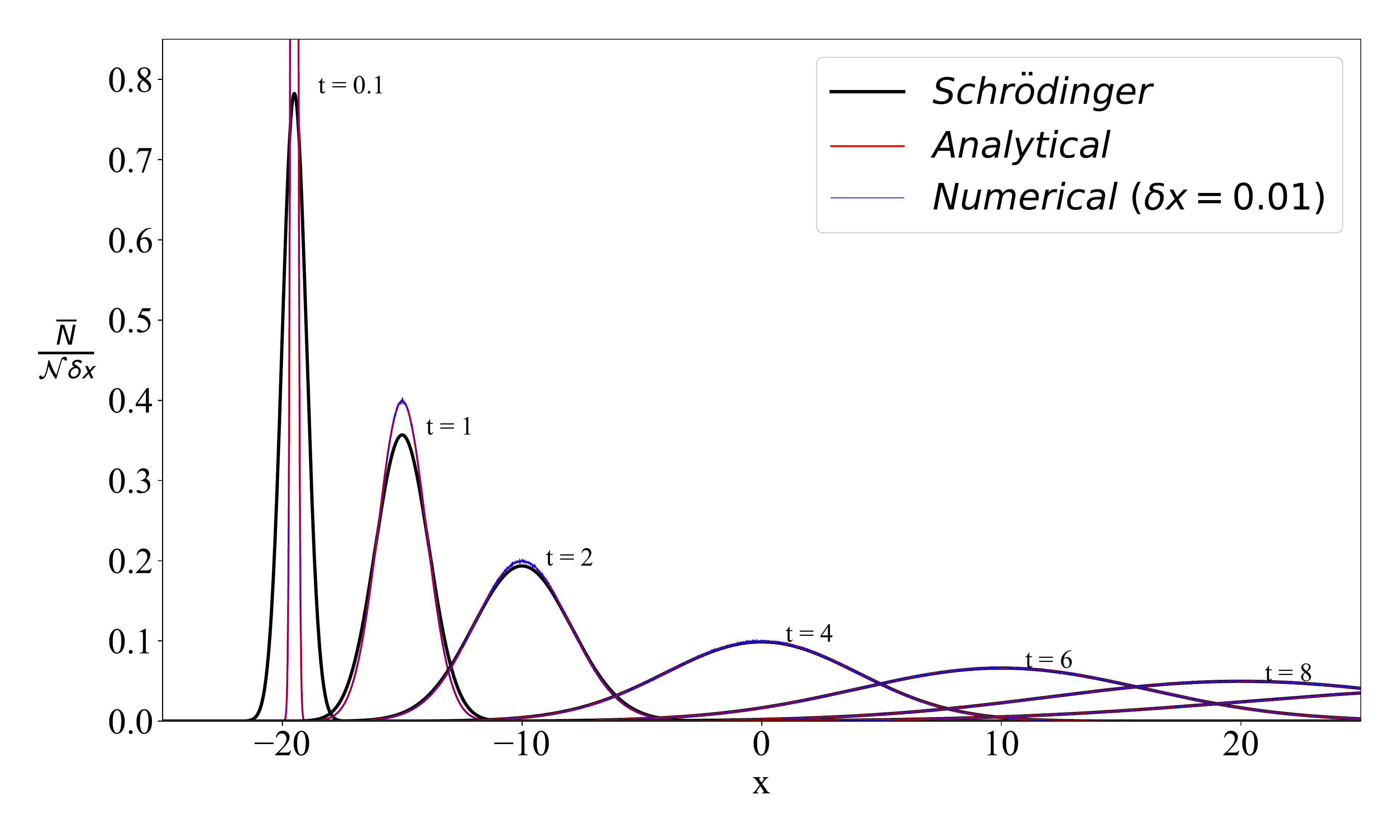}
\caption{Free-particle Gaussian distributions with average velocity ${\rm v}_o=5$ at times $t=0.1,1,2,4,6,8,10$. 
{
Here, velocities ${\rm v}_o $ are expressed in units of $\sigma_{v_o}$, spatial scales $x$ in units of  $\hbar/(m\sigma_{v_o})$, and times $t$ in units of $\hbar/(m\sigma_{v_o}^2)$ (see main text for details).
}
Black lines: solution of the Schr\"{o}dinger equation (eq.~\ref{Psisquare}). Blue lines: the particle distribution $\overline{N}(x;t)/({\cal N} \delta x)$ obtained with ${\cal N}=10,000,000$ particles and detection bin sizes $\delta x=0.01$ in our numerical experiment.
Red lines: simple analytic approximation (eq.~\ref{Noneparticle}). The three distributions are practically indistinguishable beyond time $t\gsim 2$.}
\label{onewavepacket}
\end{figure*}

We perform a numerical experiment with non-interacting individual classical particles and a special type of detector as described above. Our goal is to obtain the solution of the Schr\"{o}dinger equation
\begin{equation}
{\it i}\hbar \frac{\partial\Psi_{\rm Schr.}}{\partial t}=
-\frac{\hbar^2}{2m}\frac{\partial^2\Psi_{\rm Schr.}}{\partial x^2}
\label{Schr1}
\end{equation}
for a free particle. The solution of the Schr\"{o}dinger equation that describes a one-dimensional free Gaussian wavepacket that originates at position $x_o$ with spatial dispersion $\sigma_{x_o}$ and moves to the right with average velocity ${\rm v}_o$ is given by
\begin{equation}
\Psi_{{\rm Schr.}1}(x;t)=
\frac{{\rm e}^{-\left(\frac{m{\rm v}_o\sigma_{x_o}}{\hbar}\right)^2}}{\sqrt[4]{2\pi}\sqrt{\sigma_{x_o}+{\it i}\frac{\hbar t}{2m \sigma_{x_o}}}}\ 
{\rm e}^{\frac{\left(\frac{2m{\rm v}_o \sigma_{x_o}^2}{\hbar}+{\it i}(x-x_o)\right)^2}{4\sigma_{x_o}^2+\frac{2{\it i}\hbar t}{m}}}
\ ,
\label{single_wavepacket}
\end{equation}
with
\begin{equation}
|\Psi_{{\rm Schr.}1}|^2(x;t)=\frac{1}{\sqrt{2\pi}\sqrt{\sigma_{x_o}^2+\frac{\hbar^2 t^2}{4m^2 \sigma_{x_o}^2}}}\ 
{\rm e}^{-\frac{(x-x_o-{\rm v}_o t)^2}{2\left(\sigma_{x_o}^2+\frac{\hbar^2 t^2}{4m^2 \sigma_{x_o}^2}\right)}}
\ .
\label{Psisquare}
\end{equation}
The distribution in eq.~(\ref{Psisquare}) is Gaussian centered around $x=x_o+{\rm v}_o t$ with dispersion $\sigma_x=\sqrt{\sigma_{x_o}^2+\hbar^2 t^2/(4m^2\sigma_{x_o}^2)}$ that grows with time as $\sigma_x(t)\approx \hbar t/(2m \sigma_{x_o})$ for large times $t$. 

Let us now specify the initial conditions of our numerical experiment that are found (after several trials) to give the best agrement with the solution of the Schr\"{o}dinger equation.
Let's consider a distribution of particles all of which originate at $x=x_o$ at $t=0$ with common initial phases $\phi_o$ and random initial velocities $v_o$ that follow a Gaussian distribution
\begin{equation}
P_v(v_o)=\frac{1}{\sqrt{2\pi}}{\rm e}^{
-\frac{(v_o-{\rm v}_o)^2}{2 \sigma_{v_o}^2}}\ .
\label{Pv}
\end{equation}
These are free particles that move with constant velocity equal to their initial velocity $v_o$ obtained from the above Gaussian distribution. We follow ${\cal N}=10,000,000$ of these particles and collect their phases $\phi_j$ in detection bins of spatial width $\delta x$ along their trajectories. Their phases are then processed according to eq.~(\ref{Psi}). In what follows, units are chosen such that $\hbar/m=\sigma_{v_o}=1$. In those units, spatial scales are measured in units of $\hbar/(m\sigma_{v_o})$.
Figure~\ref{onewavepacket} shows the distributions of $\overline{N}(x;t)/({\cal N} \delta x)$ (blue lines), and compares them with the solution of Schr\"{o}dinger's equation for $\sigma_{x_o}=1/(2\sigma_{v_o})=1/2$ in eq.~(\ref{Psisquare}) in our units (black line). It is important that $\sigma_{x_o}\cdot \sigma_{v_o}=1/2$ as expected in a minimum spread quantum wavepacket. The two distributions are practically indistinguishable at all times $t\gsim 2$. In this experiment the detection spatial bin size is equal to $\delta x=0.01$. As seen in figure~\ref{growth} below, as $\delta x$ increases above that value, the addition of different particle phases in each detection bin results in the gradual reduction of the number of particles  recorded by our detectors. This effect is clearly artificial. Particles do not disappear, but our detectors will undercount them if their spatial bin size is too large.

One can rederive eq.~(\ref{single_wavepacket}) analytically for large times using our model of freely moving particles carrying a phase. In our units, $\sigma_{x_o}=1/2$ and $\hbar/m=1$, therefore, in the limit $t\gg 1$,
\begin{eqnarray}
\Psi_{{\rm Schr.}1}&=&\frac{\sqrt[4]{2/\pi}}{\sqrt{1+2it}}{\rm e}^{
-\frac{(x-x_o-{\rm v}_ot)^2}{1+4t^2}(1-2it)+i{\rm v}_o\left(x-x_o-\frac{{\rm v}_ot}{2}\right)}\nonumber\\
&\approx&
\frac{{\rm e}^{-i\frac{\pi}{4}}}{\sqrt[4]{2\pi}\sqrt{t}}\ {\rm e}^{
-\frac{(x-x_o-{\rm v}_ot)^2}{4t^2}}{\rm e}^{i\frac{(x-x_o)^2}{2t}}{\rm e}^{-i\frac{\pi}{4}}\ .
\label{single_wavepacketgg1}
\end{eqnarray}
As assumed, particles move freely, i.e. they conserve their initial velocity $v_o$. Thus, if a particle reaches position $x$ at time $t$, this implies that it indeed traveled with velocity
\begin{equation}
v_o=\frac{x-x_o}{t}
\label{v}
\end{equation}
from its position of origin at $x=x_o$. According to eq.~(\ref{phi}), its phase at that position in time will be equal to
\begin{equation}
\phi=\frac{m}{2\hbar}v_o(x-x_o)+\phi_o=\frac{(x-x_o)^2}{2t}+\phi_o\ .
\end{equation}
The number $N$ of particles reaching the detection interval $\delta x$ around detection position $x$ will thus be equal to
\begin{equation}
N=P_x(x) \delta x=P_v(v_o)\frac{\partial v_o}{\partial x} \delta x
=\frac{\delta x}{\sqrt{2\pi}}\frac{1}{t}{\rm e}^{
-\frac{(v_o-{\rm v}_o)^2}{2}}\ .
\label{Px}
\end{equation}
We have defined here the particle distribution in space $P_x(x)$. From eq.~(\ref{v}), $\partial v_o/\partial x=1/t$. Finally, according to eqs.~(\ref{Psi}) and (\ref{Nrecorded}),
\begin{eqnarray}
\overline{\Psi}({\bf x};t)&\equiv&
\frac{1}{\sqrt{N}}
\sum\nolimits_{j=1}^{N} {\rm e}^{i\phi}= \sqrt{N}{\rm e}^{i\phi}\nonumber\\
&=& \frac{\sqrt{\delta x}}{\sqrt[4]{2\pi}\sqrt{t}}\ {\rm e}^{
-\frac{(x-x_o-{\rm v}_o t)^2}{4t^2}}{\rm e}^{i\frac{(x-x_o)^2}{2t}}{\rm e}^{i\phi_o}\ ,\nonumber\\
\label{Psifree1}\\
\overline{N}({\bf x};t)&\equiv& |\overline{\Psi}|^2(x;t)\nonumber\\
&=& \frac{\delta x}{\sqrt{2\pi}t}\ {\rm e}^{
-\frac{(x-x_o-{\rm v}_o t)^2}{2t^2}}\ .
\label{Noneparticle}
\end{eqnarray}
Eq.~(\ref{Psifree1}) is the same as eq.~(\ref{single_wavepacketgg1}) for $\phi_o=-\pi/4$, modulo a constant factor of $\sqrt{\delta x}$ (red line in figure~\ref{onewavepacket}). Note also that in this case without wavepacket interference, $\overline{N}=N$.

The success of reproducing the solution of the Schr\"{o}dinger equation with freely moving independent classical particles with a simple Gaussian distribution of velocities allows us to also obtain the solution for two oppositely moving Gaussian wavepackets. As will be seen next, `quantum-like' interference patterns are obtained in that case.

\section{Two oppositely moving free-particle Gaussian wavepackets}

\begin{figure*}
 \centering
 \includegraphics[width=1.05\textwidth]{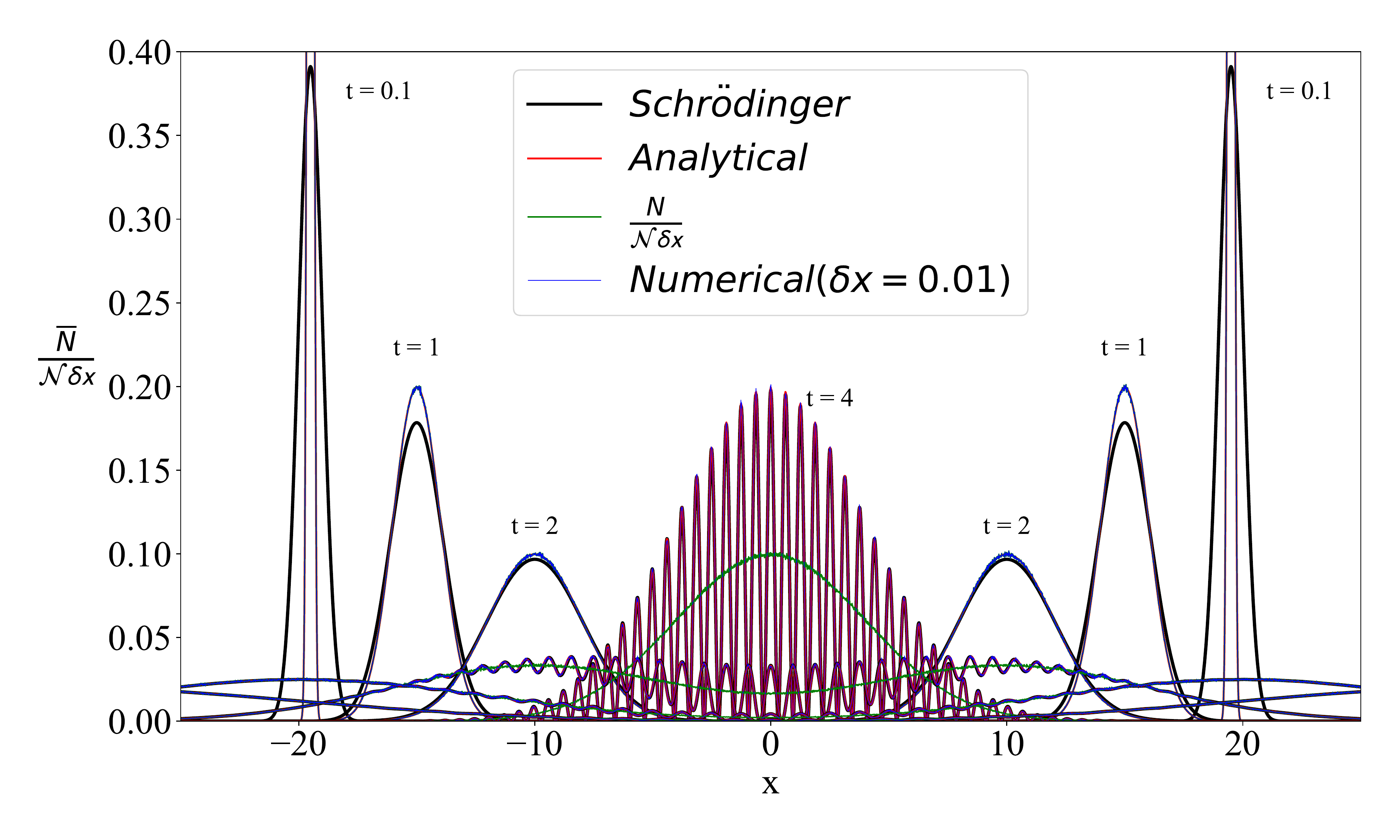}
\caption{Two free-particle Gaussian distributions moving against each other with average velocities $\pm{\rm v}_o=5$ at times $t=0.1,1,2,4,6,8,10$. Line colors as in figure~\ref{onewavepacket}.
Analytic approximation according to eq.~(\ref{Nfree}). Once again, the three distributions are practically indistinguishable beyond time $t\gsim 2$. 
Green line: the particle distribution $N(x;t)/({\cal N} \delta x)$ obtained numerically {\it without} the phase information. We see clearly two Gaussian distributions passing each other without interference.}
\label{twowavepackets}
\end{figure*}
\begin{figure*}
 \centering
 \includegraphics[width=1.1\textwidth]{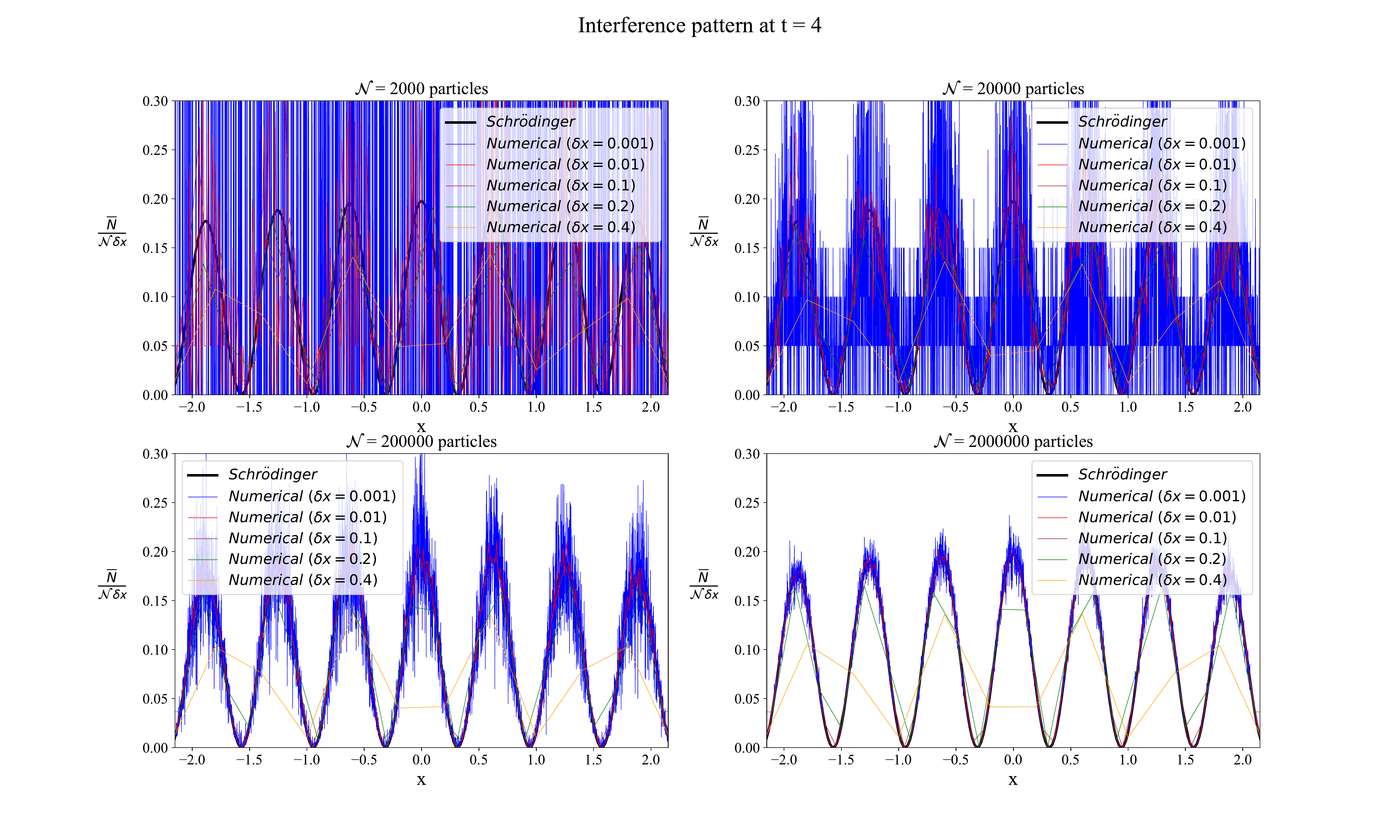}
\caption{Development of the interference pattern shown in figure~\ref{twowavepackets} at time $t=4$ for various values of the detection bin size $\delta x=0.001, 0.01, 0.1, 0.2, 0.4$ as the number ${\cal N}$ of particles/experiments increases. Black line: solution of the Schr\"{o}dinger equation at time $t=4$ as in figure~\ref{twowavepackets}. The red line corresponds to the bin size implemented to obtain figure~\ref{twowavepackets}, namely $\delta x=0.01$. As $\delta x\rightarrow 0$, it takes a very long time for the interference pattern to develop clearly. For $\delta x\gsim 0.1$, particles are undercounted because of phase mixing in each detection bin. The interference pattern disappears altogether when $\delta x$ is too large.}
\label{growth}
\end{figure*}

The quantum mechanical solution for two oppositely moving `entangled' wavepackets originating at positions $x_{1o}$ and $x_{2o}$ is given by
\begin{eqnarray}
\Psi_{\rm Schr.}&=&\frac{1}{\sqrt{2}}\left(\Psi_{{\rm Schr.}1}+\Psi_{{\rm Schr.}2}\right)\ ,
\label{two_particles}
\end{eqnarray}
where,
\begin{equation}
\Psi_{{\rm Schr.}1} = \frac{\sqrt[4]{2/\pi}}{\sqrt{1+2{\it i}t}}\ {\rm e}^{
-\frac{(x_{1o}-x-{\rm v}_o t)^2}{1+2{\it i}t}+{\it i{\rm v}_o}\left(x_{1o}-x-\frac{{\rm v}_o t}{2}\right)}\ ,\nonumber
\end{equation}
\begin{equation}
\Psi_{{\rm Schr.}2} = \frac{\sqrt[4]{2/\pi}}{\sqrt{1+2{\it i}t}}\ {\rm e}^{
-\frac{(x_{2o}-x-{\rm v}_o t)^2}{1+2{\it i}t}+{\it i{\rm v}_o}\left(x_{2o}-x-\frac{{\rm v}_o t}{2}\right)}\ .\nonumber\\
\label{Psileft}
\end{equation}
Let us now repeat our numerical experiment in direct analogy to the discussion in the previous section. As before, 10,000,000 particles are sent to the right from position $x_{1o}$ with velocity $v_{1o}$, average velocity $\langle v_{1o}\rangle=+{\rm v}_o$ and velocity dispersion $\sigma_{v_{1o}}=1$, and another 10,000,000 particles to the left from position $x_{2o}$ with velocity $v_{2o}$, average velocity $\langle v_{2o}\rangle =-{\rm v}_o$ and velocity dispersion $\sigma_{v_{2o}}=1$. The respective Gaussian velocity distributions are
\begin{eqnarray}
P_v(v_{1o})&=&\frac{1}{2}\frac{1}{\sqrt{2\pi}}{\rm e}^{
-\frac{(v_{1o}-{\rm v}_o)^2}{2}}\ ,\\
P_v(v_{2o})&=&\frac{1}{2}\frac{1}{\sqrt{2\pi}}{\rm e}^{
-\frac{(v_{2o}+{\rm v}_o)^2}{2}}\ .
\label{Pv2}
\end{eqnarray}
Notice the extra factor of $1/2$ in the above equations because we consider an entangled quantum state and not two independent quantum wavepackets.
As before, the particle phases are collected  in detection bins of spatial width $\delta x$ and calculate their distribution $\overline{N}(x;t)/({\cal N} \delta x)$ according to eq.~(\ref{Nrecorded}). There are now ${\cal N}=20,000,000$, and two particle directions, hence $K=2$. Our results for $\delta x=0.01$ are shown in figure~\ref{twowavepackets}. Once again, our results (blue line) are almost indistinguishable from the solution of the Schr\"{o}dinger equation (black line). For comparison, the distributions of particles $N(x;t)/(\delta x{\cal N})$ obtained numerically {\it without} their phase information are also plotted in figure~\ref{twowavepackets}  (green line). This shows two independent Gaussian wavepackets that move independently one through the other without interference. This clearly confirms that the phase information and our special detector counting are crucial for the manifestation of `quantum-like' interference. 

An interesting element of our numerical approach is that it takes some time for the quantum mechanical distribution to appear in our detector. 
{
Notice that it is in principle possible to determine how fast the interference pattern is developed particle-by-particle in a quantum interference experiment (e.g.~\cite{8}). In fact, this is an important parameter that may differentiate classical interpretations of quantum mechanics (e.g.~\cite{jin2010corpuscular}) from actual quantum mechanics. In other words, it is not enough that a particular theory reproduces the particle distribution in a certain quantum system. It is equally important that it does so at the same rate as an actual physical experiment of the quantum system that collects particles one-by-one (see the Appendix).
}
Figure~\ref{growth} shows how the interference pattern develops gradually as the number of particles (or equivalently the number of individual experiments) grows. The buildup time clearly depends on the size of the detection bins $\delta x$. If $\delta x$ is too small, it will take an exceedingly large time to collect enough particles at all detection positions for the interference pattern to develop clearly. On the other hand, as argued in the previous section, if $\delta x$ is very large, one will artificially record less particles  because of random phase cancelations in each detector bin. If $\delta x$ is too large, the interference pattern will disappear. 

As before, one can also obtain an analytical approximation for our result. At time $t$, particles from the left reach position $x$ with velocity
\begin{equation}
v_{1o}=\frac{x-x_{1o}}{t}\ ,
\label{v1}
\end{equation}
while particles from the right reach position $x$ with velocity
\begin{equation}
v_{2o}=\frac{x-x_{2o}}{t}\ .
\label{v2}
\end{equation}
Their corresponding numbers at a detection interval $\delta x$ around detection position $x$ are
\begin{eqnarray}
N_1 & = & P_v(v_{1o})\frac{\partial v_{1o}}{\partial x}\delta x=\frac{\delta x}{2\sqrt{2\pi}t}{\rm e}^{-\frac{(v_{1o}-{\rm v}_o)^2}{2}}\ ,
\label{N1}\\
N_2 & = & P_v(v_{2o})\frac{\partial v_{2o}}{\partial x}\delta x=\frac{\delta x}{2\sqrt{2\pi}t}{\rm e}
^{-\frac{(v_{2o}+{\rm v}_o)^2}{2}}\ .
\label{N2}
\end{eqnarray}
Their corresponding phases at the point of their detection at spacetime position $(x;t)$ are
\begin{eqnarray}
\phi_1 & = & \frac{(x-x_{1o})^2}{2t}+\phi_o\  ,\\
\phi_2 & = & \frac{(x-x_{2o})^2}{2t}+\phi_o\ .
\end{eqnarray}
Once again, according to eq.~(\ref{Psi})\ ,
\begin{eqnarray}
\overline{\Psi}({\bf x};t)&\equiv&
\frac{1}{\sqrt{N_1}}
\sum\nolimits_{j=1}^{N_1} {\rm e}^{i\phi_1}+
\frac{1}{\sqrt{N_2}}
\sum\nolimits_{j=1}^{N_2} {\rm e}^{i\phi_2}\nonumber\\
&=& \sqrt{N_1}{\rm e}^{i\phi_1}+\sqrt{N_2}{\rm e}^{i\phi_2}\nonumber\\
&=& \frac{\sqrt{\delta x}\ {\rm e}^{i\phi_o}}{\sqrt[4]{2\pi}\sqrt{2t}}\left[{\rm e}^{
-\frac{(x-x_{1o}-{\rm v}_o t)^2}{4t^2}}{\rm e}^{i\frac{(x-x_{1o})^2}{2t}}
\right.\nonumber\\
&& 
\left. 
\ \ \ \ \ \ \ \ \ \ \ +\ {\rm e}^{
-\frac{(x-x_{2o}+{\rm v}_o t)^2}{4t^2}}{\rm e}^{i\frac{(x-x_{2o})^2}{2t}}\right] \nonumber \\
\label{Psifree}
\end{eqnarray}
which is the same as eq.~(\ref{two_particles}) for $t\gg 1$ and $\phi_o=-\pi/4$, modulo a constant factor of $\sqrt{\delta x}$. 
Finally,
\begin{equation}
\overline{N}({\bf x};t)\equiv |\overline{\Psi}|^2({\bf x};t)\ .
\label{Nfree}
\end{equation}
Once again, the agreement with the solution of the Schr\"{o}dinger equation and with the result of our numerical experiment is excellent (red line in figure~\ref{twowavepackets}).

\section{Free-particle Gaussian wavepacket reflecting off a wall}

\begin{figure*}
 \centering
 \includegraphics[width=1.1\textwidth]{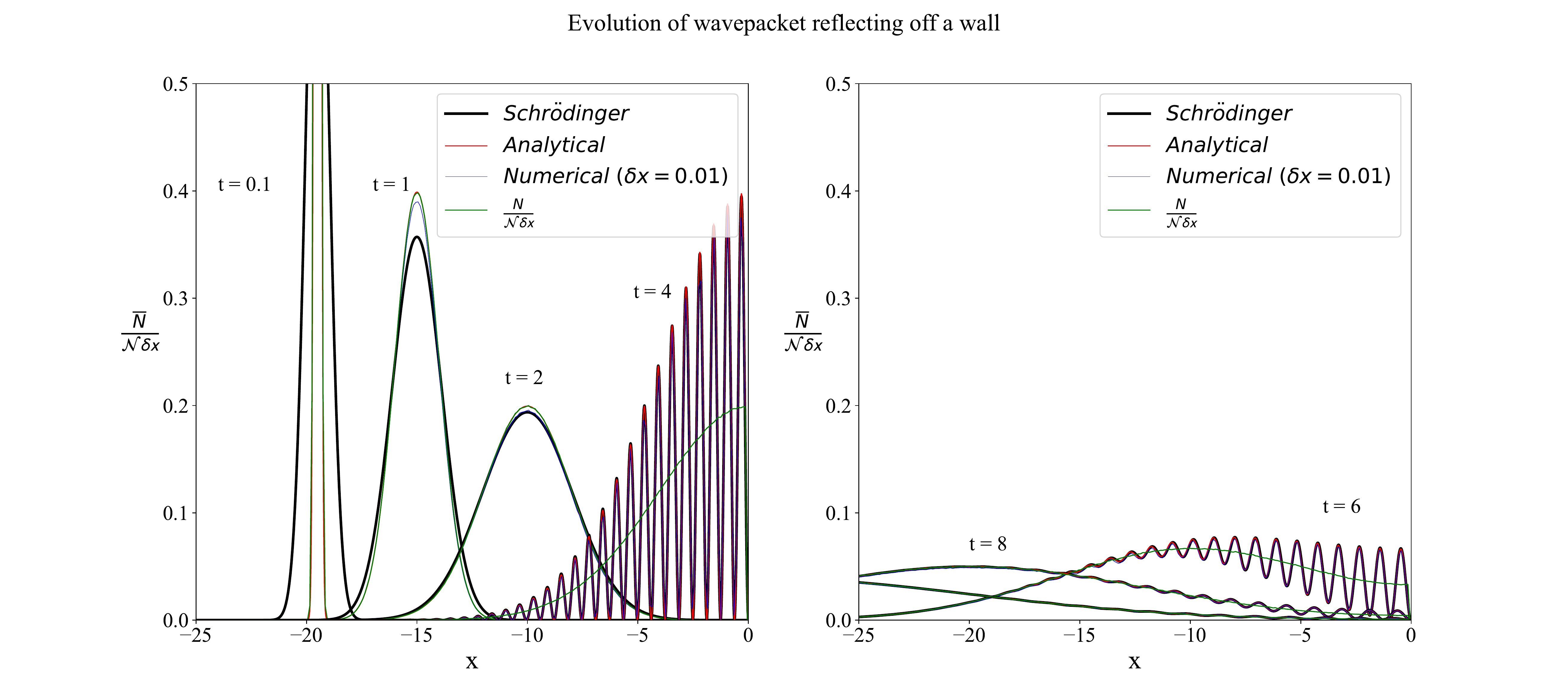}
\caption{One free-particle Gaussian distribution with average velocity ${\rm v}_o=5$ reflecting off a wall at $x=0$. Left plot: shown times $t=0.1,1,2,4$. Right plot: shown times $t=6,8,10$.
Black, blue, red line colors as in figure~\ref{onewavepacket}.
Analytic approximation according to eq.~(\ref{Nwall}). Once again, the three distributions are practically indistinguishable beyond time $t\gsim 2$. Green line: the particle distribution $N(x;t)/({\cal N} \delta x)$ {\it without} the phase information, thus also without interference.
}
\label{wall}
\end{figure*}

We finally consider an infinitely high wall at position $x=0$. Particles arrive from their initial position $x_o<0$ at position $x<0$ at time $t$ from the left with velocity and phase
\begin{eqnarray}
v_1&=&\frac{x-x_o}{t}\\
\phi_1&=&\frac{(x-x_o)^2}{2t}+\phi_o
\end{eqnarray}
respectively. Particles will also reach that position from the right after reflection from the wall with velocity and phase
\begin{eqnarray}
v_2&=&\frac{-x_o-x}{t}\\
\phi_2&=&\frac{(-x_o-x)^2}{2t}+\phi_o+\pi
\end{eqnarray}
respectively. Notice that an extra jump of $\pi$ is introduced in the phase of the reflected particle. This helps achieve zero particle density at the position of the wall as is predicted by the quantum wavefunction. In both cases, $\partial v_{1,2}/\partial x=1/t$, thus, according to eqs.~(\ref{N1}) and (\ref{N2}), their corresponding numbers at a detection position $x$ are
\begin{eqnarray}
N_1 & = & P_v(v_1)\frac{\partial v_1}{\partial x}\delta x=\frac{\delta x}{\sqrt{2\pi}t}{\rm e}^{-\frac{(v_1-{\rm v}_o)^2}{2}}\ ,\\
N_2 & = & P_v(v_2)\frac{\partial v_2}{\partial x}\delta x=
      \frac{\delta x}{\sqrt{2\pi}t}{\rm e}
^{-\frac{(v_2+{\rm v}_o)^2}{2}} \ .
\end{eqnarray}
and
\begin{eqnarray}
\overline{\Psi}(x<0;t)&=&\sqrt{N_1}{\rm e}^{i\phi_1}+\sqrt{N_2}{\rm e}^{i\phi_2}\nonumber\\
&=& \frac{\sqrt{\delta x}\ {\rm e}^{i\phi_o}}{\sqrt[4]{2\pi}\sqrt{t}}\left[{\rm e}^{
-\frac{(x-x_o-{\rm v}_o t)^2}{4t^2}}{\rm e}^{i\frac{(x-x_o)^2}{2t}}
\right.\nonumber\\
&& 
\left. 
\ \ \ \ \ \ \ \ \ \ \ -\ {\rm e}^{
-\frac{(x+x_o+{\rm v}_o t)^2}{4t^2}}{\rm e}^{i\frac{(x+x_o)^2}{2t}}\right] \ ,
\nonumber\\
\label{Psi1}\\
\overline{N}(x<0;t)&=&|\overline{\Psi}|^2(x<0;t)\ .
\label{Nwall}
\end{eqnarray}
Figure~\ref{wall} shows the excellent agreement between the numerical solution of the Schr\"{o}dinger equation
(black line), the numerical experiment of collecting 10,000,000 particles in detection bins of size $\delta x=0.01$ (blue line), and the analytic approximation of eq.~(\ref{Nwall}) (red line). In this particular example, there is no analytic solution of the Schr\"{o}dinger equation.

{
\section{Discussion}

The fact that this method yields results that are practically indistinguishable from those obtained by means of the Schr\"{o}dinger equation is not surprising, at least for free particles. The reason is that Ehrenfest’s theorem predicts that the position average, i.e., the expectation value of the position operator will follow the classical Newton’s laws. The quantum mechanical `fuzziness' in the standard Schr\"{o}dinger theory comes from the phase of the wave function. However, this phase does not evolve separately from the magnitude because their respective equations are coupled. In this paper  agreement with quantum mechanics is achieved not by means of such coupled equations but by assigning a phase and initializing a space-time distribution, each particle obtaining an extra phase due to different initial conditions. Naturally, if there is an initial uncertainty, it will propagate and evolve in time. The existence of the phase is absolutely necessary. 

}

{

As a continuation to the project, we would like to consider classical particles moving inside a potential (e.g. a square barrier or well, the Coulomb potential in the hydrogen atom, etc.). We are not interested in approximating quantum mechanical states as in the semi-classical (SC) IVR method \cite{Miller2001} where particles only follow classical trajectories. In order to achieve perfect agreement with quantum mechanics we will need to allow particles to also follow classically `forbidden' trajectories. One possibility may be to implement our methodology of phase collection and particle counting in the presence of stochastic interactions with a `background' as in Nelson's stochastic mechanics \cite{nelson2012review}. 

}

We would also like to discuss the physical meaning of our special type of detector. It will help our discussion if the analogy with photons is made. According to classical electromagnetism, an electromagnetic wave in vacuum is characterized by its electric field, its direction, and its frequency $\nu\equiv c/\lambda$ ($\lambda$ is here the wavelength of the radiation). These are the most fundamental quantities from which one can calculate the Poynting flux of energy and the electromagnetic energy density. Quantum mechanics offers a different picture of the electromagnetic wave in terms of individual photons. If one were to reconcile the wave and particle pictures without invoquing non-local interactions between photons, the flux of energy may be viewed as the flow of $N$ photons per unit time each carrying energy $\epsilon$ at the speed of light along the direction of the Poynting vector. Non-interacting photons, however, cannot carry the information of the electric field which is fundamental for the manifestation of wave behavior (e.g. refraction, diffraction, interference, etc.). The reason is that the collection of $N$ indistinguishable photons carries $N$ times the energy $\epsilon$ of one photon, but the amplitude of the electric field of the corresponding classical wave is proportional to $\sqrt{N \epsilon}$. Taking the square root of $N$ is obviously a collective operation that cannot be carried as information by non-interacting individual particles. It is interesting that eqs.~(\ref{Psi}) and (\ref{Nrecorded}) operate exactly as in electromagnetism: add first the electric vectors, and from them calculate the energy density carried by the electromagnetic field. This analogy with electromagnetism makes us hopeful that our way of adding particle phases according to eq.~(\ref{Psi}) may indeed be implemented in nature. 

{

We propose a simple physical experiment that will test whether physical detectors operate as proposed. All that is needed is a single-photon detector and a low intensity double-slit laser interference pattern. The single-photon detector will be placed first at the position of a maximum of the interference pattern. Then,
\begin{enumerate}
\item The detector will be turned on and we will wait till a single photon is detected.
\item The detector will then be turned off and on, expecting that it thus loses all memory of previous photon detections, and in particular all information about their phases \footnote{This expectation must be confirmed. We may have to wait for some longer time interval before the detector is turned on again.}.
\item The above procedure will be automated and will be repeated untill a large number of photons is recorded (e.g. 1,000 photons).
\item The time $t_{\rm max}$ it takes to complete that part of the experiment will be measured.
\item The detector will then be moved to the position of the nearby minimum and steps 1 to 3 will be repeated. The time $t_{\rm min}$ it takes to complete that second part of the experiment will also be measured.
\end{enumerate}
If $t_{\rm min}\gg t_{\rm max}$ that will imply that photons `know' via the quantum mechanical wavefunction how often to arrive on particular positions of the detection screen, and indeed populate preferentially the positions of maxima in the interference pattern. If on the other hand $t_{\rm min}$ turns out to be comparable to $t_{\rm max}$ that will imply that photons reach positions of maxima and minima at roughly the same frequency as is expected for classical particles. This would prove that the interference pattern is due to the memory of the detector at each detection position, and if that memory is somehow erased (e.g. by turning the detector off and on after each single-photon detection), the intereference pattern disappears.

}

{

\section{Conclusion}

}

{

We have shown that `quantum-like' interference patterns may be obtained with a classical detector that acts in the particular way expressed by eq.~(\ref{Nrecorded}). Our current model applies only to free particles. It is acknowledged that our mathematical formulation is related to the IVR method of the quantum computational community. The focus of our work, however, is not on the pragmatic application of the method itself. We are interested in the role of physical detectors on the emergence of quantum interference patterns. 
A tentative conclusion of the present work is that, if there are macroscopic detectors operating as proposed, then interference phenomena may be explained classically, and quantum mechanical behavior may be due to our detectors. This hypothesis (which may lead to an alternative interpretation of quantum mechanics) can be tested with a simple laser interference experiment.

}

\vspace{0.5cm}

\paragraph*{\bf Acknowledgements:}
We acknowledge fruitful discussions with Pr. Christos Efthymiopoulos. All numerical calculations shown in the main part of the paper were performed in Python 3 by FZ.

\section*{Appendix}

We would like here to briefly discuss a numerical interference experiment with massless photons \cite{PaperI}. Let's assume that individual photons of wavelength $\lambda$ carry an evolving phase 
\begin{equation}
\phi \equiv \int \frac{2\pi {\rm d}l}{\lambda}=\frac{2\pi l}{\lambda}
\ ,
\label{phiphoton}
\end{equation} 
where, once again, the ${\rm d}l$ integration is taken along the photons' trajectory from their point of origin to the finite region of spacetime where and when their detection takes place. Eq.~(\ref{phiphoton}) differs from eq.~(\ref{phi}) that is implemented for massive particles in the present work by a factor of two. Nevertheless, if eq.~(\ref{phi}) is rewritten as
\begin{equation}
\phi = 2\pi \int \frac{\frac{1}{2} mv^2{\rm d}t}{h}=
2\pi \int \frac{\epsilon_m{\rm d}t}{h}\ ,
\end{equation}
where $\epsilon_m$ is the particle kinetic energy, and  if eq.~(\ref{phiphoton}) is rewritten as
\begin{equation}
\phi = 2\pi \int \frac{c{\rm d}t}{c/\nu}=2\pi \int \frac{h\nu{\rm d}t}{h}=
2\pi \int \frac{\epsilon_{ph}{\rm d}t}{h}\ ,
\end{equation}
where $\epsilon_{ph}\equiv h\nu$ is the photon energy, the two expressions become one and the same.

\begin{figure}[!h]
 \centering
 \includegraphics[width=0.8\textwidth]{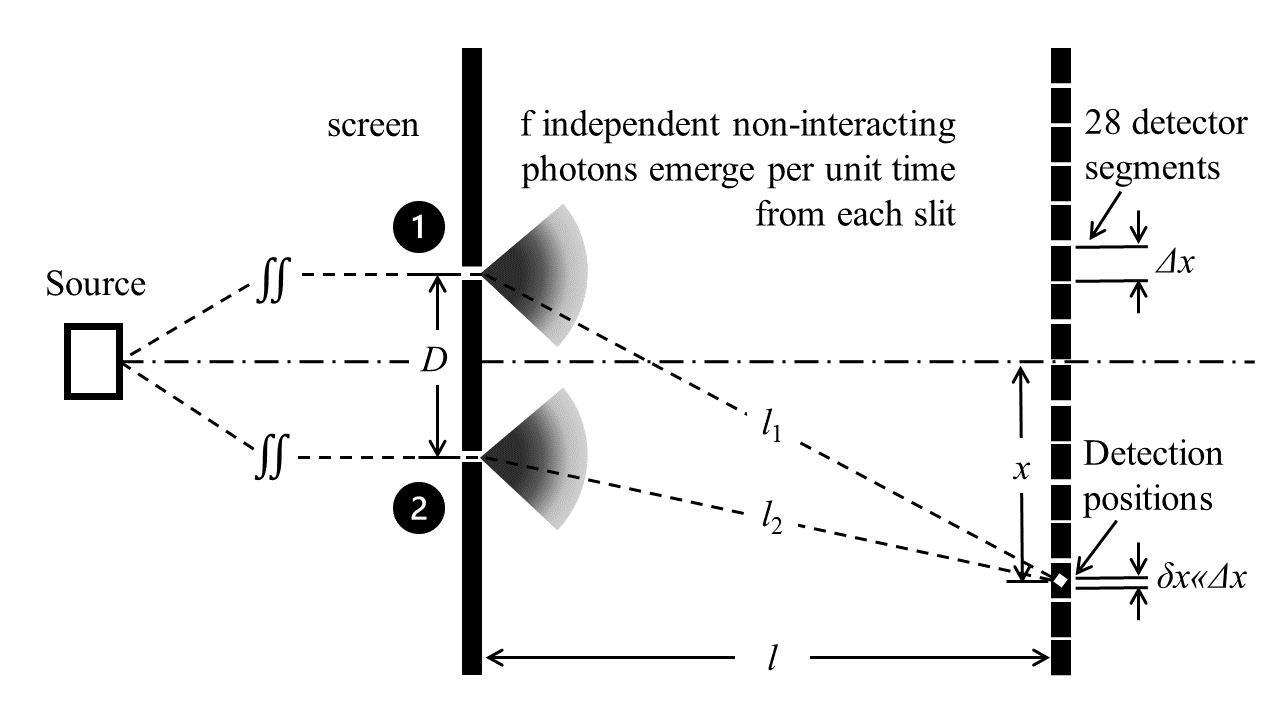}
\caption{Schematic of a double slit experiment with $f$ individual non-interacting photons emitted horizontally per unit time from each slit. $l\gg D$.}
\label{doubleslit}
\end{figure}

\begin{figure}[!h]
 \centering
 \includegraphics[width=0.8\textwidth]{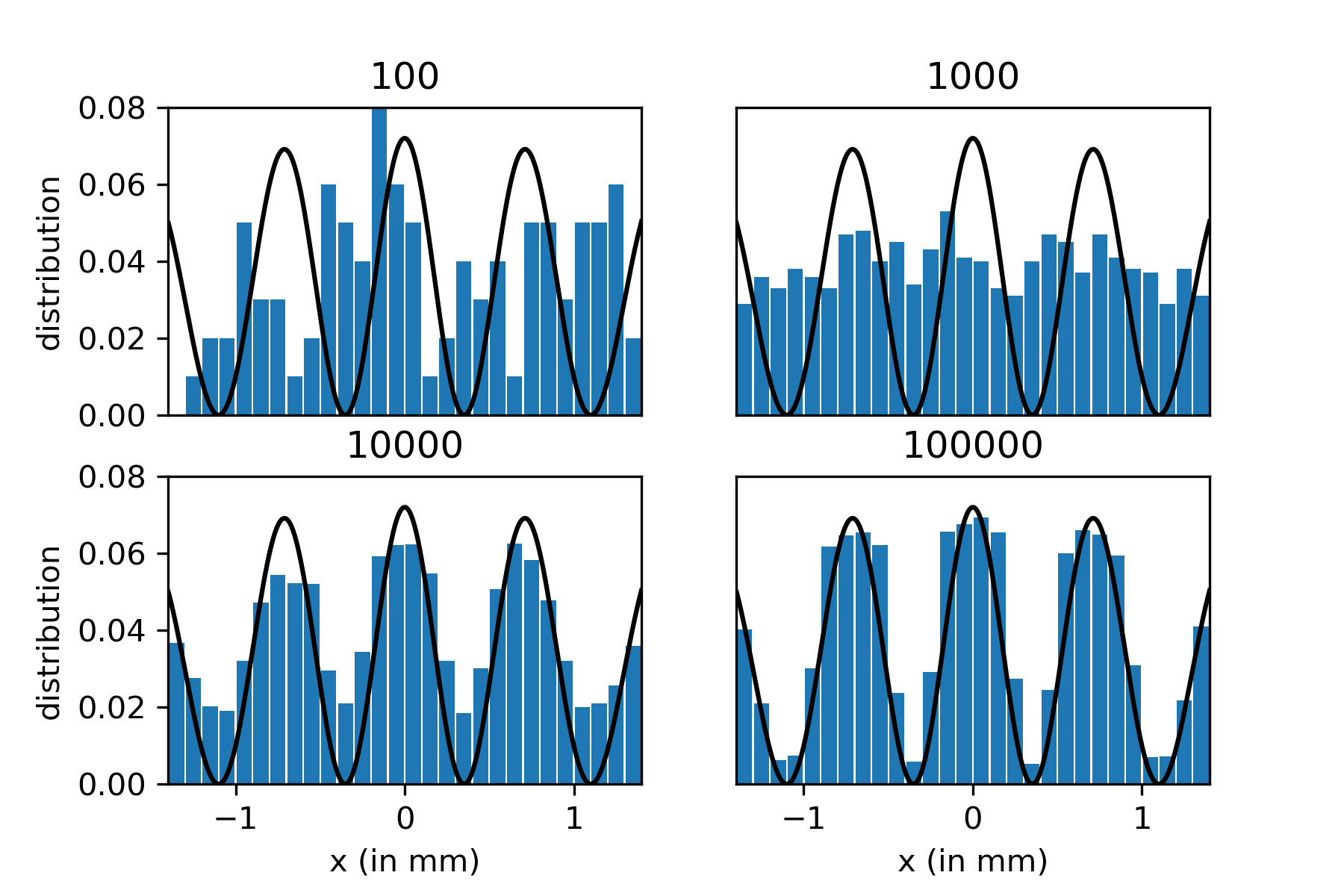}
\caption{Gradual buildup of the double-slit interference pattern for 100, 1,000, 10,000 and 100,000 photon detections respectively, analogous to figure~\ref{growth}. Shown are the distributions of photons along the 28 $100\mu$m wide segments of the detector in the actual experiment of \cite{8}. Detection pixels have width $\delta x=1\mu{\rm m}\sim \lambda$. Solid line: analytic solution.}
\label{doubleslit2}
\end{figure}

Using eq.~(\ref{phiphoton}) and the same type of detector described
by eqs.~(\ref{Psi}-\ref{Nrecorded}), it is possible to numerically reproduce the main results of the double-slit experiment performed by \cite{8}. We have considered individual photons of wavelength $\lambda=842$~nm that emerge horizontally from a certain source and reach a screen with two parallel perpendicular slits separated by a distance $D$ (figure~\ref{doubleslit}). Photons emerge at a rate of $f$ photons per unit time from each slit at the speed of light, they are distributed isotropically on the horizontal plane, and then hit a horizontal detector some distance $l\gg D$ behind the screen. The detector consists of an array of segments of width $\Delta x=100$~$\mu$m where individual photons are collected and their phases are processed according to eq.~(\ref{Psi}). Each detection position has a width $\delta x=1$~$\mu$m at positions $x$ from the detector midpoint. Figure~\ref{doubleslit2} shows the gradual buildup of the double-slit interference pattern for 100, 1,000, 10,000 and 100,000 photon detections respectively, similarly to figure~\ref{growth}. In order to obtain an interference pattern from the collection of a reasonable number of photons, the width $\delta x$ of the detection pixels must be neither too small, neither too large (see \cite{PaperI} for details). It is also assumed that there are no other external perturbations acting on the photons, thus all photons move along straight lines during their flight from the source to their eventual detection at the detector. 

\end{document}